\documentclass{emulateapj}
\usepackage{bm}

\newcommand{\bt}{\mbox{\boldmath{$\theta$}}}
\newcommand{\bb}{\mbox{\boldmath{$\beta$}}}

\usepackage{natbib}
\begin{document}

\title{Fold Lens Flux Anomalies: A Geometric Approach}

\author{David M. Goldberg, Mary K. Chessey, Wendy B. Harris, Gordon
  T. Richards}

\affil{Department of Physics, Drexel University, Philadelphia, PA
  19104}

\email{goldberg@drexel.edu}

\begin{abstract}
  We develop a new approach for studying flux anomalies in
  quadruply-imaged fold lens systems.  We show that in the absence of
  substructure, microlensing, or differential absorption, the expected
  flux ratios of a fold pair can be tightly constrained using only
  geometric arguments.  We apply this technique to 11 known quadruple
  lens systems in the radio and infrared, and compare our estimates to
  the Monte Carlo based results of Keeton, Gaudi, and Petters (2005).
  We show that a robust estimate for a flux ratio from a smoothly
  varying potential can be found, and at long wavelengths those lenses
  deviating from from this ratio almost certainly contain significant
  substructure.
\end{abstract}

\maketitle

\section{Introduction}

To date, at least forty-eight multiply lensed quasars with three or
more images have been discovered \citep[e.g. CASTLES\footnote{See
  http://cfa-www.harvard.edu/castles},][]{ Hewitt/Turner:1992,
  Inada/Oguri:2005, More/McKean:2009}. The great advantage to studying
multiply imaged systems is that symmetries of the lensing galaxy can
be understood without a detailed mass reconstruction \citep{PLW}.

\begin{figure}[h!]
\centerline{\includegraphics[width=7in,angle=0]{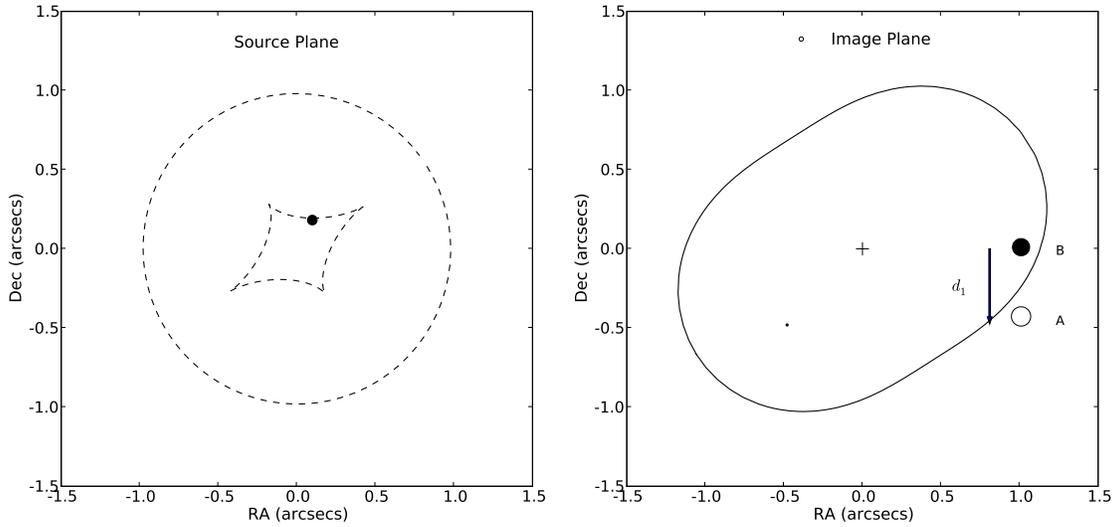}}
\caption{A typical fold lens. {\em Left:} Source plane with caustic
  curves.  {\em Right:} Image plane with critical curves.  Images with
  negative magnification (parity) are denoted with filled circles, while
  images with positive magnification (parity) are open
  circles. Throughout the paper (and in most papers on fold lenses)
  Image A is the positive parity fold image, and Image B is the
  negative parity fold image.  Following KGP, $d_1$ is the distance
  separating A and B. }
\label{fg:fold}
\end{figure}

Of particular interest are the so-called ``fold'' lenses
~\citep[][hereafter KGP]{Keeton/Gaudi:2005} in which two images lie on
opposite sides of the tangential critical curve in the image plane,
while in the source plane, the source lies near an edge of a
tangential caustic.  We illustrate the geometry of a fold lens in
Fig.~\ref{fg:fold}.

As a source gets arbitrarily close to the caustic it can be shown from
purely analytic arguments that magnification of the two images should
be equal and opposite \citep{BlandfordNarayan,PLW,Schneider1992}.
Thus, we would naively expect the fold relation:
\begin{equation}
R_{fold}\equiv\frac{f_A-f_B}{f_A+f_B}\ ,
\end{equation}
to be zero, where $f_{A,B}$ are the fluxes of the individual images in
the same band.  Our convention is that images with a negative parity
still have a positive flux.

Observationally, however
\citep[e.g.][]{Pooley/Blackburne:2006,Pooley/Blackburne:2007,Keeton/Burles:2006}
there is often a significant flux anomaly between two images.  there
has been significant discussion on the nature of the fold flux
anomalies
\citep{Mao/Schneider:1998,Kochanek/Dalal,Congdon/Keeton:2005}.  For
optical lenses, two of the most common explanations include:
microlensing from stars in the lensing galaxy
\citep{Koopsmans/Bruyn:2000,Metcalf/Madau:2001,Schechter/Wambsganss:2002,
  Keeton:2003,Chartas/Eracleous:2004,Morgan/Kochanek:2006,
  Anguita/Faure:2008} and differential reddening by dust
\citep{Lawrence/Elston:1995}. These causes of flux anomalies are
expected to be highly wavelength-dependent, however.  For example,
differential absorption will strongly affect optical photometry, but
will have limited or no effect in the infrared (IR) or radio.  Though
microlensing applies achromatically to the lensed image, it is only a
significant effect when the Einstein radius of the lensing star is
similar to or larger than the angular size of an emission region in a
particular waveband.  Typically, Einstein radii of individual stars at
cosmological distances (i.e., in the lensing galaxy) will be on the
order of microarcseconds which corresponds to a typical angular size
of quasar optical emission regions.  However, radio emission,
especially from radio lobes, is generally much more extended.  As a
result, microlensing is less likely to cause significant flux anomalies
in the radio.  

While the current work focuses exclusively on fold lenses, future
geometric analysis may shed light on ``cusp lenses''
\citep{Keeton/Gaudi:2003,PLW}, in which a source image near the corner
of a caustic produces 3 clustered images in the foreground plane.  As
with folds, cusps are naively expected to obey a simple flux ratio
relation in which the brightest image precisely equals the sum of the
fluxes of the dimmer two.  As with folds, significant deviations from
this expectation have been observed.

Even if we confine the discussion to observations at long wavelength,
fold flux anomalies don't disappear.  This is likely attributable to
small-scale variations in the lens galaxy potential
\citep{Congdon/Keeton:2005, Mao/Schneider:1998}.  To that end, most
workers have focused on generating semi-analytic models which closely
fit the observed image positions and fluxes.  They propose adding
explicit substructure to their models as a kludge to correct the
fluxes.  There has been enormous effort expended trying to model
galaxy lenses explicitly \citep{Munoz/Kochanek:2001} using {\tt
  lensmodel} \citep{Keeton:2001} and other software.  However, for
many of these lenses \citep{Evans/Witt:2003, Mao/Schneider:1998} no
simple analytic model will suffice.  Some researchers
\citep{Evans/Witt:2003,Congdon/Keeton:2005} decompose lensing
potentials into an orthonormal basis sets or multipole moments and
show that any set of observational constraints may be fit with a
complex enough expansion of the potential.  It is noteworthy, however,
that all of these authors stress that not all of these models provide
viable explanations of the flux anomalies.

In reality there is a great deal of information about the local
lensing field which can be gleaned geometrically.  That is, using only
the positions of the observed lensed images, we will derive a
semi-analytic approach to smooth lens flux anomalies.  A particular
system can then be analyzed without recourse to complicated models.

In this paper, we will develop a semi-analytic geometric approach to
understanding fold lenses.  This approach is intended as a first step toward a
more general theory involving measurement of galaxy lens substructure.
Our approach is as follows.  In \S\ref{sec:theory}, we derive a
geometry-based expansion of a smooth potential, and discuss how
observables can be used to uniquely predict a flux anomaly.  In
\S\ref{sec:simulations} we test our semi-analytic results with
simulated galaxy lenses, and show that our approach allows us to
identify lenses with significant substructure.  In \S\ref{sec:data},
we describe an observational set of fold lenses in the radio and IR
regime, and in \S\ref{sec:results}, we apply our sample.  Finally, in
\S\ref{sec:future} we discuss future prospects.

\section{Theory}

\label{sec:theory}

\subsection{Notation}

It is useful to give a bit of background regarding the notation we'll
be using throughout.  As is the normal practice, we will use the
angular vector, $\bb$ to describe (non-observable) positions in the
background plane in the absence of lensing.  Likewise, we use $\bt$
for the observed position(s) of the lensed image.  They are related
via:
\begin{equation}
\beta_i = \theta_i - \alpha_i(\bt)
\label{eq:lensing}
\end{equation}
where $i=1,2$ used for the principle directions, and the displacement
vector is defined as:
\begin{equation}
\alpha_i=\psi_{,i}\ .
\end{equation}
The subscripted index represents a single angular derivative in the
$\theta_i$ direction.  Since we will be performing multiple
derivatives, subsequent derivatives of the potential will omit the
comma in order to reduce clutter.

The potential is generated via the two-dimensional Poisson equation:
\begin{equation}
\nabla^2\psi=2\kappa
\end{equation}
where $\kappa$, the convergence, is the dimensionless surface density
of the galaxy.  

Likewise, the Jacobian of the source position generates
two shear terms:
\begin{eqnarray}
\gamma_1&=&\frac{\psi_{11}-\psi_{22}}{2}\\
\gamma_2&=&\psi_{12}\\
\end{eqnarray}
where
\begin{equation}
\gamma^2=\gamma_1^2+\gamma_2^2
\end{equation}

This yields an inverse magnification field of:
\begin{equation}
\mu^{-1}=(1-\kappa)^2-\gamma^2
\label{eq:mu_inv}
\end{equation}
where along ``critical curves'' of the lens, $\mu^{-1}=0$.

\subsection{Fold Flux Anomalies}

KGP derived an analytic expression for the expected fold relation
arising from a smooth potential.  Their expression involved the Taylor
expansion of the potential around a point on the critical curve.  In
order to estimate the fold relation, $R_{fold}$, in the simplest form, they
performed a rotation around the center of the lens such that the fold
images are oriented vertically from one another, with the positive
parity image (``A'') below the negative parity one (``B'').  In
\cite{saasfees}, the interested reader can find some very helpful figures to
illustrate image parity.  In Fig.~\ref{fg:orient}, we show our rotated
coordinate system (along with a few other angles to be used later).

By definition, along the critical curve, equation~(\ref{eq:mu_inv})
equals zero. The rotated coordinates were specifically selected such
that $\psi_{12}=0$, and $\psi_{22}=1$.  This rotation can be performed
without a loss of generality.  Likewise, for our derivation, we
further allow a reflection such that the rotated images appear in the
positive-x half of the plane.

We may parameterize the fold ratio with a simple form:
\begin{equation}
R_{fold} = A_{fold} d_1
\end{equation}
where throughout, we will refer to $A_{fold}$ as the ``anomaly
parameter'', and $R_{fold}$ as the ``fold relation.''  The
anomaly parameter is introduced because it can be shown to be a
constant in the limit of small separations.  KGP showed that the
anomaly parameter may be expressed as:
\begin{equation}
  A_{fold} = \frac{3\psi_{122}^2 - 3\psi_{112}\psi_{222}+\psi_{2222}(1
    - \psi_{11})}{6\psi_{222}(1 - \psi_{11})}
\label{eq:afold}
\end{equation}
and $d_1$ is the distance between the two images in the observed
plane.  We note that this expression contains one term ($\psi_{11}$)
related to the local shear field, many (the third derivatives) related
to the local flexion field \citep[e.g.,][]{Goldberg/Bacon,Bacon}, and 1
term related to the fourth derivative, which is normally not
considered at all in lensing analysis.

\subsection{Simplified Analytic Models}

\begin{figure}
\centerline{\includegraphics[width=5in]{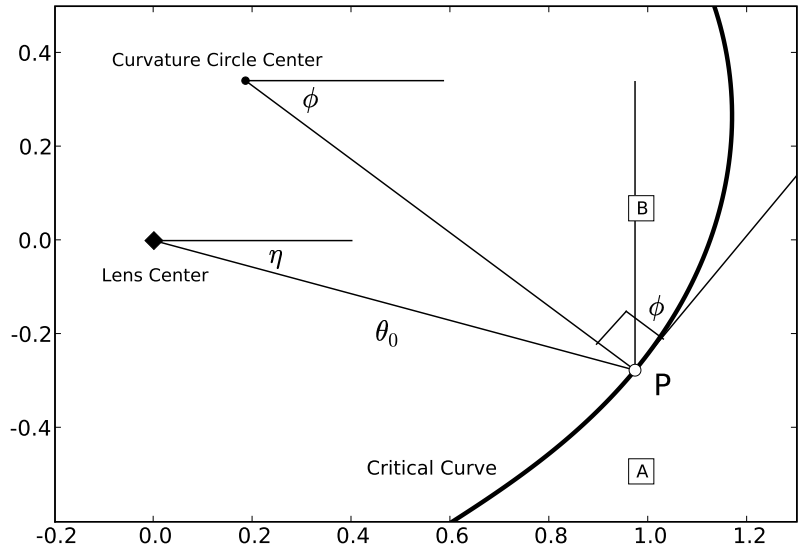}}
\caption{The relative orientation of the fold images (A,B), the lens center,
  and the critical curve in our model.  The point, ${\cal P}$,
  represents the position on the critical curve where the two images
  would meet if the source were moved closer to the caustic in the
  source plane.  It is also the point around which all derivatives of
  the potential are defined.
}
\label{fg:orient}
\end{figure}

KGP model a number of observed multiple lensed systems using a wide
range of elliptical isothermal models, and show that there is a small
reasonable range of flux anomalies that might be expected for known
fold systems.  However, for smooth potential models, simulations
aren't necessary.  For this paper, we will define a ``smooth
potential'' very strictly, but will subsequently show that even
non-smooth potentials can be adequately fit.  For our derivation, we
assume:
\begin{enumerate}
\item That the potential must be expressible as a circularly
symmetric potential, plus an external shear:
\begin{equation}
\psi^{(tot)}(\bt)=\psi(\theta)+\gamma^e \theta^2\cos(2\nu)
\label{eq:psidef}
\end{equation}
where $\gamma^e$ is the magnitude of the external shear, and $\nu$ is the
angle between the induced shear and the radial vector.   In our
rotated coordinate frame, this is essentially the angle that the
best-fit critical curve ellipse makes with the horizontal.

\item That the shape of the critical curve is largely independent of a
  detailed model of the potential.  This is due in part to the fact
  that the critical curve must thread between the 4 images in a quad
  system.  This is one of the main reasons that the image positions of
  quad systems may be fit very generically while fluxes are more
  complicated to fit.

  In other words, we will generate a simple lens model using an
  isothermal sphere plus external shear based on positions only.  We
  argue that the shape of the critical curve will not vary
  significantly from other models, an assumption we will test in
  \S\ref{sec:simulations}.
\item All of our expansion will be around an as yet unknown point,
  ${\cal P}$, which lies along the critical curve.  This is the same
  point used in equation~(\ref{eq:afold}) to determine the $A_{fold}$
  parameter (and related to that, the local derivatives of the
  potential).  The vector between the center of the lens and ${\cal
    P}$ makes an angle, $\eta$ with respect to the horizontal (as
  illustrated in Fig.\ref{fg:orient}).  This angle is assumed to be
  small.  For the sample discussed in \S\ref{sec:data}, the maximum
  $\eta$ is 26 degrees with an average over the sample of only 12
  degrees.  
\end{enumerate}

With full generality, we can apply local geometric properties of the
critical curve to simplify equation~(\ref{eq:afold}).  First, we note
that at ${\cal P}$:
\begin{eqnarray}
  \nabla(\mu^{-1})&&= \nonumber\\
&&\left[-\psi_{111}(1-\psi_{22})-\psi_{122}(1-\psi_{11})-
    2\psi_{12}\psi_{112}\right]\hat{i}
  \nonumber \\ 
  +&&\left[-\psi_{112}(1-\psi_{22})- 
\psi_{222}(1-\psi_{11})-2\psi_{12}\psi_{122}\right]\hat{j}\nonumber \\ 
  =&&-(1-\psi_{11})(\psi_{122}\hat{i}+\psi_{222}\hat{j})
\end{eqnarray}
where the simplification can be found by applying the constraints on
the second derivatives from the choice of rotation.

Since the gradient is perpendicular to the critical curve, if the
curve makes an angle, $\phi$, with vertical (as shown in
Fig.\ref{fg:orient}) then:
\begin{equation}
\tan\phi=-\frac{\psi_{222}}{\psi_{122}}
\end{equation}
again, with complete generality.

We now Taylor expand the circular component of the potential around ${\cal
  P}$, such that:
\begin{equation}
\psi=(\theta-\theta_0)\psi'+\frac{1}{2}(\theta-\theta_0)^2\psi''+
\frac{1}{6}(\theta-\theta_0)^3\psi'''+\frac{1}{24}(\theta-\theta_0)^4\psi''''+...
\label{eq:taylor}
\end{equation}
where ${\cal P}$ is a distance $\theta_0$ from the center of the lens
and primes represent radial derivatives of the circular component of
the potential.  We can then expand equation~(\ref{eq:taylor})
explicitly, such that $\theta_x=\theta_0\cos\eta$ and
$\theta_y=\theta_0\sin\eta$.  

For an assumed value of $\eta$, we can thus compute all higher
derivatives of the potential.  While the Taylor expansion does not
include the external shear component, it should be noted that the only
term in equation~(\ref{eq:afold}) which will be affected by the
external shear is $\psi_{11}$.  All 3rd derivatives and higher will
include only the circular component of the potential.

Thus, we may expand the ratio as a series in $\sin\eta$
\begin{eqnarray}
\frac{\psi_{222}}{\psi_{122}}&=&3\sin\eta+
\frac{1}{2}\left(
\frac{4\psi'''\theta_0^2-15\psi''\theta_0+15\psi'}{\psi'-\theta_0\psi''}\right)
\sin^3\eta+... \nonumber \\ 
&=&-\tan\phi
\label{eq:taylorpotential}
\end{eqnarray}
For an isothermal circular component, the quantity in the parentheses
reduces to 15, while for a point mass, it becomes 4.  In any event,
under under the assumption that $\eta$ is small, only the first term
matters, and we get a relationship which is largely independent of
radial profile.  For small $\eta$:
\begin{equation}
\phi\simeq -3\eta
\label{eq:etasimp}
\end{equation}

Thus, for a fiducial model of the critical curve near the fold, a
unique position, ${\cal P}$, can be found.  As a first step in the
procedure, consider the critical curve near the midpoint of the two
fold images.  About the midpoint, measure the local curvature.  Using
standard trigonometry, this arc uniquely defines a circle in the lens
plane with a center at coordinates $(x_c,y_c)$, with radius of
curvature, $r$, from which some tedious algebra yields:
\begin{equation}
\sin\eta\simeq
\frac{y_c-r\sin\phi}{C\sqrt{1-\frac{2y_cr\sin\phi}{C^2}}}
\label{eq:sineta}
\end{equation}
where
\begin{equation}
C\equiv \sqrt{r^2+x_c^2+y_c^2+2x_cr\cos\phi}
\end{equation}
Combined with equation~(\ref{eq:etasimp}), $\eta$ can be solved
iteratively very quickly.

Objections might be raised that application of
equation~(\ref{eq:sineta}) requires that we have a global model of the
potential.  This is true, but as we will show in
\S~\ref{sec:simulations}, a wide range of models will leave the local
shape of the critical curve largely unchanged.  Moreover, for most
lenses, a good estimate of $\eta$ can be found by simply taking the
midpoint of the two fold images.

\subsection{Semi-Analytic Fold Ratio Estimates}

Once we have an estimate of $\eta$, it is a straightforward matter to
estimate the flux anomaly parameter, $A_{fold}$ in terms of the radial
derivatives of the potential field using equation~(\ref{eq:taylor}).
Each of the terms in equation~(\ref{eq:afold}) may then be expanded
explicitly.  If we define a smooth field as a circular profile plus an
external shear, only 2nd derivatives contain any indication
of the non-circularity. 

In particular, note that to first order in $\eta$:
\begin{equation}
\psi_{22}\simeq \frac{\psi'}{\theta_0}-\gamma^e\cos(2\nu)=1
\end{equation}
where the last equality is guaranteed by the choice of coordinates.
Thus, we have:
\begin{equation}
\psi'=\theta_0\left[1+\gamma^e\cos(2\nu)\right]
\end{equation}
It should be noted that the estimate of $\gamma^e$ comes from the
fiducial radial profile of the model used to fit the critical curve.
For small values of $\gamma^e$, for example, a critical curve can be
approximated as:
\begin{equation}
\theta(\nu)=\theta_E\frac{\psi''(\theta_E)}{\left(1-\psi''(\theta_E)\right)^2}
\left(\psi''-1+2\gamma^e\cos(2\nu)\right)
\end{equation}
This naturally means that there is a degeneracy in shape such that for
fixed shape of a critical curve:
\begin{equation}
\gamma^e\propto 1-\psi''(\theta_E)\ .
\end{equation}
Henceforth, it will be assumed that $\gamma^e$ is the shear estimated
by fitting the critical curve to a Singular Isothermal Sphere plus an
external shear.  If we wish to model the curve using a different
radial profile, a correction of $\gamma^e(1-\psi'')$ must be included.

We are now ready to Taylor expand equation~(\ref{eq:afold}) into a
form that can be estimated only using direct observables, and an
assumed power-law radial profile for a circular lens.  In the flux
anomaly parameter, only $\psi_{11}$ has any dependence on the external
shear terms.  For the rest, we may easily relate all terms with a
combination of radial derivatives and trigonometric functions in
$\eta$.   For example:
\begin{equation}
\psi_{11}=\frac{\psi'({\cal P})\sin^2(\eta)+\theta_0 \psi''({\cal
    P})\cos^2(\eta)+\theta_0\gamma^e\cos(2 \nu)}{\theta_0}
\end{equation}
and similarly for higher derivatives.  Combining all terms, and
dropping all those quadratic or higher in $\gamma^e$, we get:
\begin{eqnarray}
A_{fold}&\simeq&
-\frac{\gamma^e\cos(2\nu)}{3\theta_0\eta}\left(\frac{1-\psi''(\theta_E)}{1-\psi''}\right)\nonumber
\\ 
&&-\frac{\eta}{\theta_0}
\left(\frac{1-\psi''+\frac{1}{2}\theta_0\psi'''}{1-\psi''}\right)
\label{eq:Asimp}
\end{eqnarray}

Equation~(\ref{eq:Asimp}) may seem complicated, but most terms are
immediately measurable either directly (e.g. $\theta_0$) or through a
fiducial model ($\eta$). The shear terms ($\gamma^e$, $\nu$), in
particular, can be deduced nearly uniquely from the observed galaxy
and quasar image positions.  The various radial derivatives can be
computed from various circular potential models, but assuming that
$\theta_0\simeq \theta_E$, the range of possible flux anomalies is
quite narrow.

Some examples:
\begin{enumerate}
\item Isothermal Sphere (at the Einstein radius): $\psi''=0$, $\psi'''=0$, so:
\begin{displaymath}
A_{fold}^{(SIS)}\simeq \frac{\gamma^e\cos(2\nu)}{3\theta_0 \eta}-\frac{\eta}{\theta_0}
\end{displaymath}
\item Point Source (at the Einstein radius): $\psi''\simeq -1$,
  $\psi'''\simeq As w1/\theta$:
\begin{displaymath}
  A_{fold}^{(PS)}\simeq -\frac{\gamma^e\cos(2\nu)}{3\theta_0\eta}-1.5\frac{\eta}{\theta_0}
\end{displaymath}
\end{enumerate}

Note that these are differences of less than two in the anomaly
parameter, $A_{fold}$, over a fairly wide range of potentials.  In the
next section, we will show that even if we use an incorrect global
model for a fold lens, we are still able to reproduce accurate fold
ratios for the smooth component of the system.  Finally, it will be
noted that at least one term each of the anomaly parameter models has
$\eta$ in the denominator.  Further, we have noted that $\eta$ is
necessarily a small angle.  It is precisely because of this form that
the anomaly parameter can quite large.  However, since the external
shear appears in the numerator, and that term is also typically small,
we do not find any systems in which the estimated fold relation is
divergently large.

\section{Simulations}

\label{sec:simulations}

\subsection{Smooth Model Reconstructions}

\label{sec:smooth}

As a test of the geometric fold approach, we run a number of simple
model galaxies through {\tt lensmodel}.  The source galaxy was chosen
to be in near proximity to the lens caustic, producing a fold.  While
most of the the source positions were put in ``by hand'' they were
selected to produce foreground image positions consistent with a
strict definition of a fold.  For us this means that the pair needed
to be separated by less than half the characteristic radius of the
system.  As KGP point out, however, the position along the caustic
significantly affects the expected fold relation.  Thus, for our first
set of simulations, those for a Singular Isothermal Sphere with
external shear, we've been careful to explore the fold relation for
those source along the caustic compared to those along a radius in the
source plane.

In each case, we have generated a simulated image set, including image
positions and fluxes, and have assumed a knowledge of image parities.
We have also assumed that the lensing galaxy centroid position is
known.  For the observed images, we then perform a simple lens model
fitting, assuming only image positions, and using a very simple model
of a singular isothermal sphere with external shear (regardless of the
``true'' lensing galaxy).  We then compare the resulting flux anomaly
parameter ($A_{fold}$) to the ``true'' parameter found from the image
fluxes.  \\

\subsubsection{Singular Isothermal Sphere+External Shear}

As a first test, the lens galaxy consisted of a simple Singular
Isothermal Sphere (SIS) with Einstein radius, $\theta_E=1$, with an
additional external shear, $\gamma^e=0.0.15$.  Though this is a
somewhat higher external shear than typically observed
\citep[e.g.][]{HolderSchechter}, we've selected such highly elliptical
models as an upper bound on reasonable physical systems.  Since for all
models we use an SIS+external shear model to reconstruct an estimate
of the critical curve, to some degree, this simulation is simply a
test of our algebra.

\begin{figure}[h]
\centerline{\includegraphics[width=6.0in]{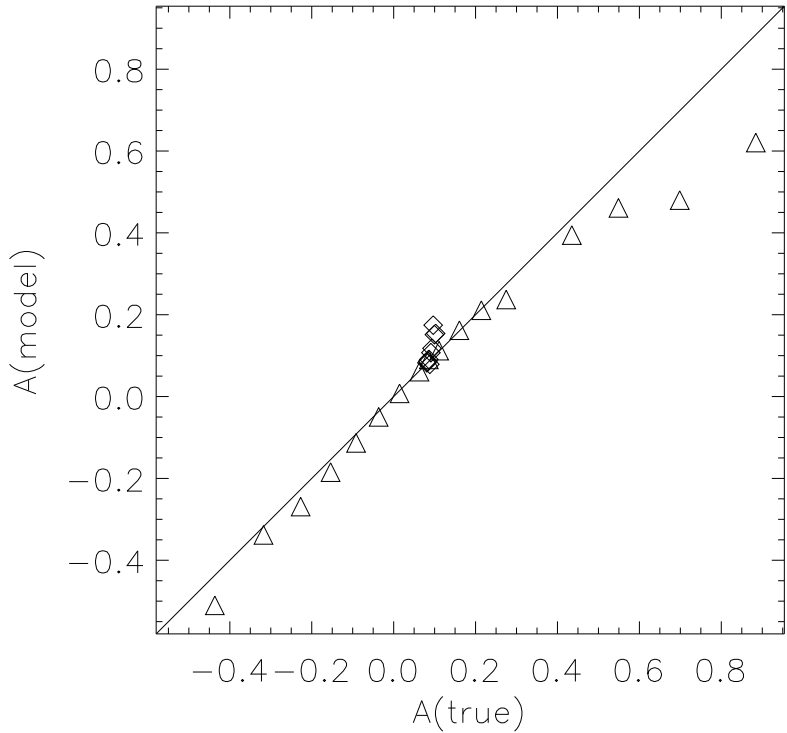}}
\caption{The measured and estimated flux anomaly parameter for a
  Singular Isothermal Sphere (SIS) lens with an external shear.  This
  simulation, and those in subsequent figures, was produced using the
  {\tt lensmodel} package.  In each case, the lens has an Einstein
  radius of $\theta_e=1$, and in this model, there is an external
  shear of $\gamma^e=0.15$.  As KGP have shown, we get dramatically
  different flux anomalies depending on where the source image lies
  along the caustic.  Triangles represent sources near the edge of a
  caustic.  We have uniformly placed 18 sources approximately 10\% the
  characteristic scale from, and along the edge.  The large flux
  ratios arise for sources nearer to the cusps.  Unsurprisingly, those
  sources near cusps more strongly resemble ``cusp lenses.''  Since
  cusp lenses are expected to have a different flux relationship (in
  which the bright image equals the sum of two dimmer ones), it is
  unsurprising that the flux anomaly between two images can be rather
  large.  Indeed, this is precisely the result found by KGP.  In every
  case, the geometric reconstruction technique produces a very good
  fit ($\langle \delta A \rangle = 0.02$, $\delta R\le 0.01$).  The 8
  diamonds represent images placed radially inward from the center of
  the caustic to within 10\% of the center of the source.  Those with
  a small ($\delta A\simeq 0.1$) deviation from prediction are those
  closest to the center, and most closely resemble quad lenses, rather
  than true folds.  The diagonal line is meant as a guide and has a
  slope of 1.}
\label{fg:ASIS}
\end{figure}

In Fig.~(\ref{fg:ASIS}), we plot the measured flux anomaly parameters
against those estimated by our geometric approach. We took a set of
sources lying along the edge of the caustic, as well as another placed
along a radius.  By far most of the variation in anomaly parameter was
due to position along the caustic rather than in radius.  The mean
error in the reconstructed anomaly parameter is $\langle \delta
A\equiv A_{est}-A_{true} \rangle \le 0.02$, corresponding to an error
in the flux anomaly of $|\delta R|\le 0.01$.  To reduce clutter, we
have shortened, $A_{fold,est}$ to simply $A_{est}$, here and
elsewhere.  This is far smaller than the factor of $\sim 2$ arising
from an uncertainty in the radial profile of the circular component of
the
lens.  \\

\subsubsection{Point Source+External Shear}

\begin{figure}[h]
\centerline{\includegraphics[width=6.0in]{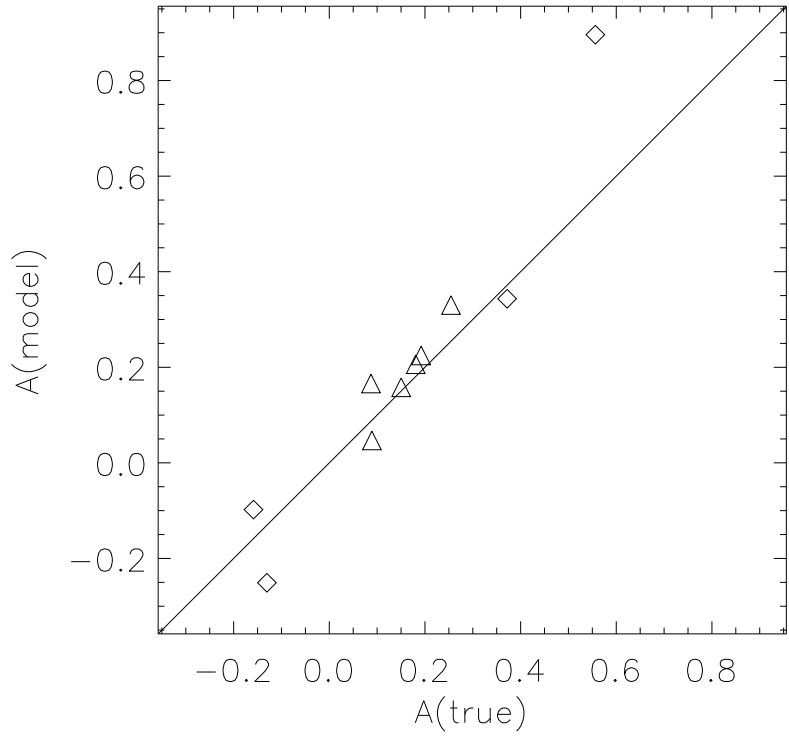}}
\caption{The measured and estimated flux anomaly parameter for a Point
  Source (PS) lens with an external shear.  In each case, the lens has
  an Einstein radius of $\theta_e=1$, and various systems were modeled
  with an external shear of $\gamma^e=0.2$ and $0.3$.  Since most of
  the variation in flux anomaly arises for variations along the edge
  of the caustic, we use a slightly different set of symbols in this
  figure and subsequently than in the previous one.  Triangles
  represent sources near the edge of a caustic, while squares
  represent sources closer to cusps.  In both cases, there is a small
  systematic error in the anomaly parameter of $\langle \delta
  A\rangle\simeq 0.08$.  This corresponds to a typical error in the
  flux anomaly of $\delta R\simeq 0.04$.  }
\label{fg:APS}
\end{figure}

As a second test, our true lens consists of a point source (again with
$\theta_E=1$) with an external shear of 0.2 or 0.3.  This is a test
that the true radial profile has little effect on the reconstructed
shape of the critical curve.  We plot the reconstructed flux anomaly
parameters against the true anomaly parameter in Fig.~\ref{fg:APS}.
Unlike in the previous test, points were selected randomly to lie near
the critical curve such that the observed images satisfy the fold
condition.  

Note that $\langle \delta A \rangle \simeq 0.03$ for sources near the
middle of folds, while $\langle \delta A \rangle \simeq 0.04$ for
sources nearer to cusps, resulting in a typical $\delta R\simeq 0.02$.
Since truly anomalous flux ratios tend to be in the neighborhood of
$R\simeq 0.5$, this
is the difference of only a few per cent.\\

\subsubsection{Singular Isothermal Ellipsoid}

\begin{figure}[h]
\centerline{\includegraphics[width=6.0in]{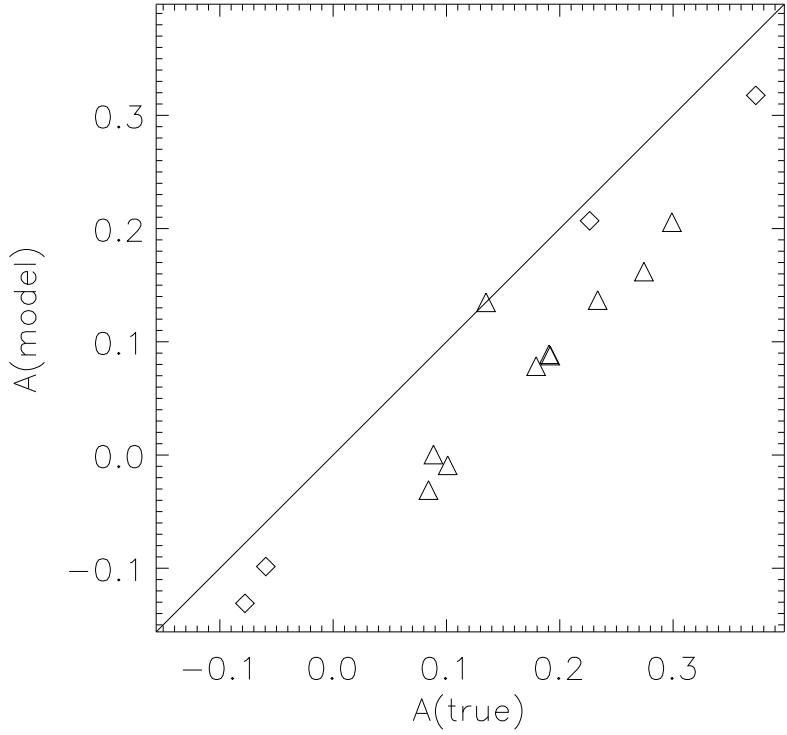}}
\caption{The measured and estimated flux anomaly parameter for a
  Singular Isothermal Ellipsoid (SIE) with an ellipticity of 0.5.  The
  fiducial lens has an Einstein radius of $\theta_e=1$.  As in the
  previous figures, triangles represent sources near the edge of a
  caustic, while squares represent sources closer to cusps.  Cuspy
  folds seem well modeled, with $\langle \delta A \rangle=-0.02$.
  However, sources further from the cusps had a larger systematic
  error, with $\langle \delta A\rangle\simeq -0.09$.  This one-sided
  offset is primarily due to a flaw in our assumptions.  Throughout,
  we've assumed that the shear only contributes in second derivatives
  of the potential.  In future work, we may relax this condition
  somewhat and allow for an intrinsic elliptical potential.}
\label{fg:ASIE}
\end{figure}

In much of our derivation, and in particular, in
equation~(\ref{eq:taylorpotential}), we made the assumption that third
derivatives of the potential contained contributions from only the
circularly symmetric part of the potential.  This is exactly true for
external shears, but not, in general, for elliptical potentials.
However, as shown by \cite{Keeton/Kochanek/Seljak}, there is an
approximate degeneracy between external shears and ellipticities, and
we will exploit this here.  As a check of whether our analytic
derivation was good enough, we have estimated flux anomalies for
simulated Singular Isothermal Ellipsoids, in this case, with a lens
ellipticity of 0.5.  As above, we've simulated fold systems and
reconstructed them using the geometric approach.  Our results are
plotted in Fig.~\ref{fg:ASIE}.

While cuspy folds are modeled extremely well, with $\langle \delta A
\rangle\simeq -0.02$, more typical folds exhibit a larger systematic
error, with $\langle \delta A\rangle \simeq -0.09$.  Of course, this
results in a systematic error in the measured flux anomaly of $|\delta
R|\simeq 0.04$ for typical fold systems, an effect smaller than even
typical observational uncertainties for many lenses.  Since an SIE
does not have as many similar properties to other mass models that
include external shear, testing our geometric method with a ``true''
elliptical lens re-fit to a simpler mass model with external shear
should result with values that give some error to the flux
anomaly. Our results prove that even big differences in the properties
of the true lens don't contribute a significant amount of error to the
flux anomaly when the assumption is made that the lens is a simple
mass model with some amount of external shear.

\subsection{The Critical Curve}

In the simulations above, we have shown that the geometric model can
be used to predict the flux anomaly of a smooth lens with good
precision, regardless of whether we use the ``correct'' model to
estimate the critical curve of the lens.  Because the critical curve
must essentially thread the observed images, we have contended that
the shape of the critical curve (and hence the values of $\eta$,
$\gamma^e$ and $\nu$ used in equation~\ref{eq:Asimp}) will largely be
independent of the smooth lens model used to produce it.

\begin{figure}[h]
\includegraphics[width=6.5in]{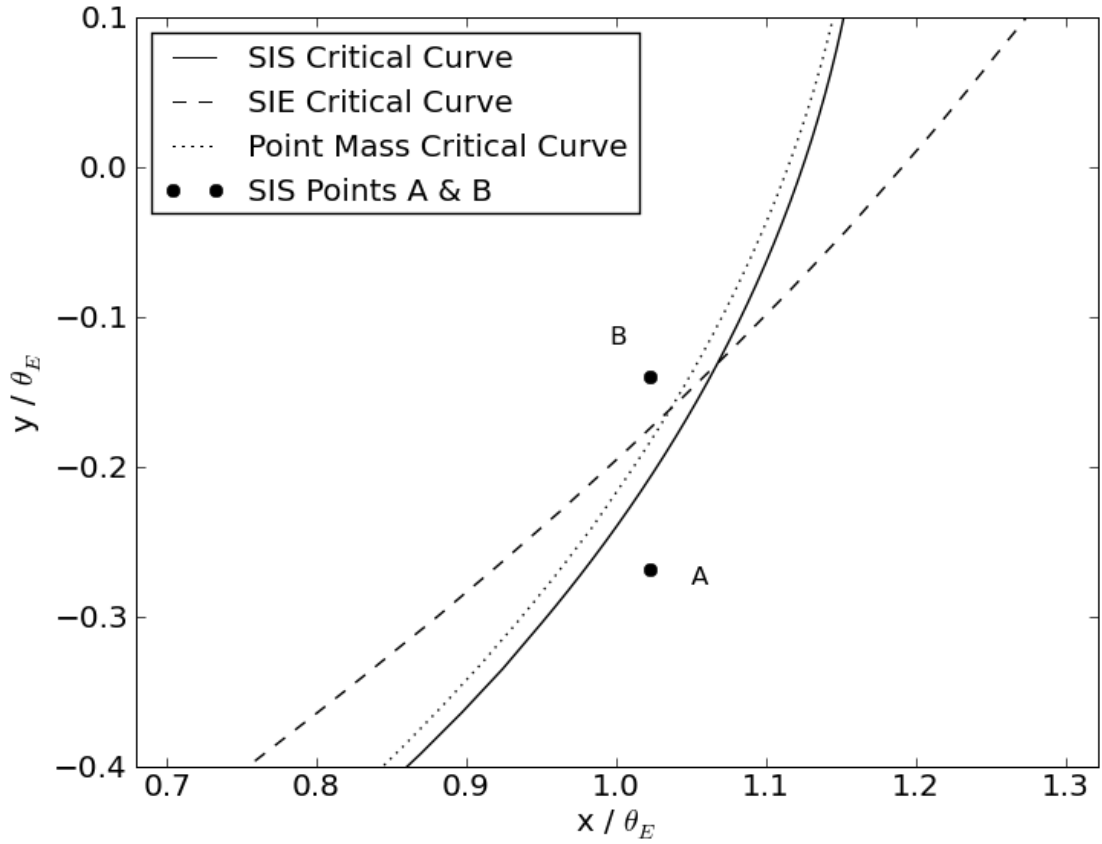}
\caption{  The fold lens images for a simulated quad system with a
  Singular Isothermal Sphere ($\theta_E=1$) and an external shear with
  $\gamma_e=0.2$.  Only the two fold images are shown in this plot,
  but the critical curves are fit to all four image positions, and the
  image centroid. Curves are fit for several models with external
  shear, as well as potentials with intrinsic ellipticity. The solid
  line critical curve represents a Singlular Isothermal Sphere with
  external shear, the dashed line represents a Singular Isothermal
  Ellipsoid and the dotted line represents a Point Source with external
  shear.}
\label{fg:comparecrit}
\end{figure}

Starting with an SIS with external shear of 0.2, we generated a fold
image pair from placing a source near the middle of the fold of the
caustic. The image positions were re-fit using a number of models: 1) A
SIS with external shear, 2) A point source
with external shear and 3) A Singular Isothermal Ellipsoids. Fig.~(\ref{fg:comparecrit})
illustrate the critical curves for the various reconstructions.

By the shape of the critical curves near the images, we have proven
that the values of $\eta$, $\gamma^e$ and $\nu$ are independent of the
mass model used, thus the values generated from the three mass models
are generally the same. For the models with external shear: $\langle
\delta \nu \equiv \nu_{SIE}-\nu_{PS} \rangle \simeq 0.002$ and
$\langle \delta \eta \equiv \eta_{SIE}-\eta_{PS} \rangle \simeq
0.020$.  What is most noticeable from this plot and the above results
is that the radii of curvature of all of the external shear models (1
and 2 in the list above) have very nearly the same radius of curvature
and orientation.  The ellipticity model (3) seems to produce
relatively different curves and it might be expected that they would
produce significantly different models for the flux anomaly parameter.
As we've seen above in \S\ref{sec:smooth}, this is not the case. This
further demonstrates that refitting image positions to a very simple
mass model, such as an Isothermal Sphere with external shear, still
provides a good estimate of the flux anomaly parameter even if the
true mass model's value of $\eta$ based on it's critical curve is not
close to the modeled values.

\subsection{A Foray into substructure}

\label{sec:substructure}

The effect of substructure in multiple-imaged quasars has been
well-studied in both simulation
\citep{Mao/Schneider:1998,Metcalf/Madau:2001,Kochanek/Dalal,
  Dobler/Keeton:2006, Williams/Foley:2008} and in a number of observed
systems
\citep{Bradac/Schneider:2002,Chiba/Minezaki:2005,Miranda/Jetzer:2007,
  More/McKean:2009}.  One of the major motivations for our geometric
approach is that we anticipate being able to unambiguously estimate
substructure within the cluster potential.  The shape of the critical
curve is largely dominated by the smooth component of the potential,
while the flux anomaly depends explicitly on higher derivatives and
thus can be significantly affected by substructure.  Moreover, we
propose that model-dependent simulations are not necessary.

As a simple proof of concept, we simulated a point mass lens with
$\theta_E=1$ and an external shear of $\gamma^e=0.2$.  We then
considered the effect on a fold lens if we placed a second point mass
lens with $\theta_E=0.2$ (a realistic 4$\%$ mass perturbation) in the
proximity of the fold.  In each case, the point mass was placed a
distance comparable to $d_1$, and subsequently the substructure mass
was rotated around the fold pair.  Even in an extreme case, the shape
of the critical curve near the fold -- and thus the positions of the
images -- is largely unaffected.  Indeed, in almost every model (save
one), the estimated flux anomaly parameter (generated geometrically)
was within $20\%$ of the anomaly parameter for the unsmoothed system.

\begin{figure}[h!]
\includegraphics[width=6.5in]{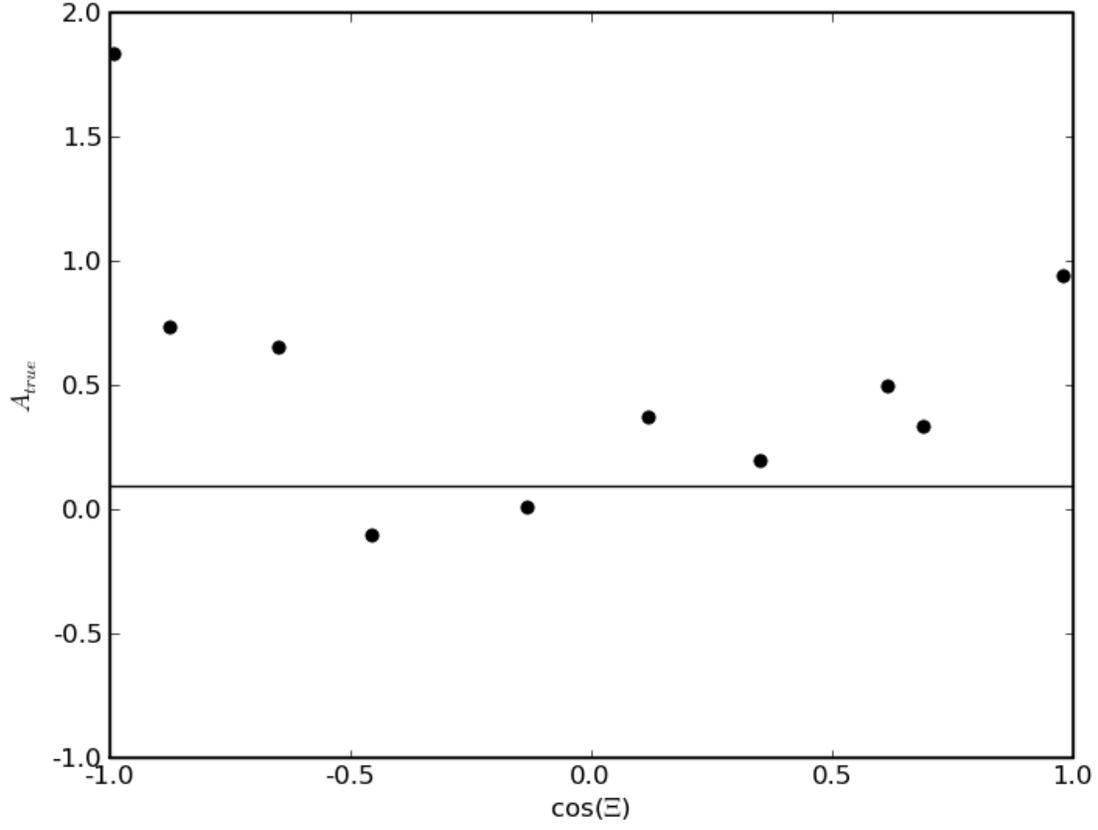}
\caption{As described in the text, the observed flux anomaly parameter
  for a variety of simulated fold lens systems.  For each, a point
  mass with external shear was placed at the origin.  In addition, a
  substructure with $\theta_E=0.2$ was placed a fixed separation
  (approximately equal to the image pair) such that the angle between
  the image separation and the secondary mass was $\Xi$. }
\label{fg:Asubfig}
\end{figure}

However, the observed flux anomaly was another matter entirely.  In
Fig.~\ref{fg:Asubfig} we show the dependence of $A$ on the angular
position of the substructure.  
Moving forward,
it is anticipated that substructure can be included in our geometric
model up to a degeneracy in mass, separation, and position angle of
the source.

\section{Observational Sample}

\label{sec:data}

We will now apply our new approach to observed fold lenses.  We have
collected eleven currently known fold lenses observed in the radio and
IR.  A summary of the observational references of the systems can be
found in Table~\ref{tab:data}.  A lens is included as a ``fold'' if it
was identified as such in its primary observation paper.  In addition,
we consider a fold pair if the separation between the two is less than
$0.5\theta_E$ (where $\theta_E$ is the best-fit circular lens profile)
and the next closest pair has separation greater than $\theta_E$.

In all the fold lenses analyzed here, we assign image ``A'' to be
outside the tangential critical curve and to have positive parity,
regardless of the designation originally given by the
investigators.  Likewise, image ``B'' is the image inside the critical
curve with negative parity. 

When information about the position of the lensing galaxy in a system
is unknown (B1555+375, B1608+656, B1933+503), we use Keeton's {\tt
  lensmodel} software to make an estimate of its location.  Otherwise,
our models are based on the positions of the observed images only.  As
discussed in the simulations section (above), the local shape of the
critical curve (the only piece of information required in our
estimates) is not strongly dependent upon choice of model.  For
consistency, we choose to use a Singular Isothermal Sphere (hereafter
SIS) with external shear for the lens.  In the case of B0128+437,
ellipticity is also included to produce a sufficiently accurate
recreation of the observed image positions.

\section{Results}

\label{sec:results}

\begin{figure}[h!]
\centerline{\includegraphics[width=5in]{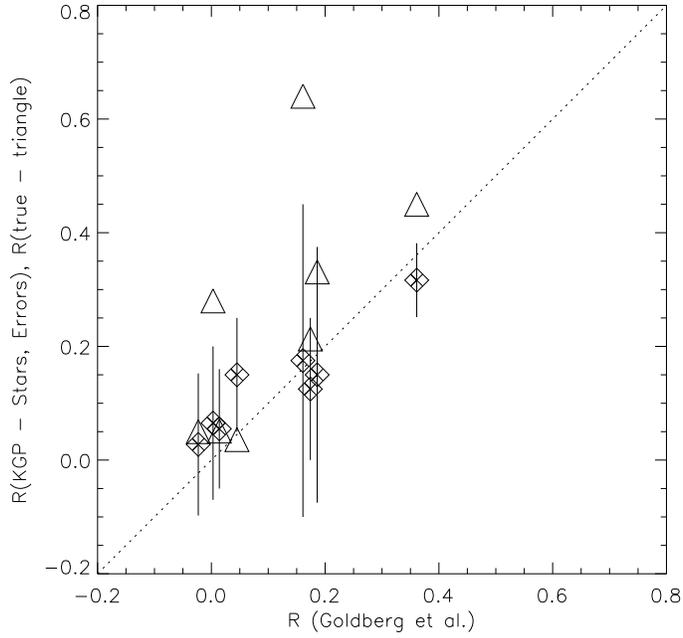}}
\caption{A comparison of our simulated Isothermal Sphere+external
  shear estimates of the fold relation (x-axis), with those
  estimated by KGP (diamonds with errorbars) from Monte Carlo
  simulation, and observed (triangles).  This sample consists of the 10
  quad systems in the radio and IR for which analysis was done by both
  groups.  Note that our results have a very close
  correspondence ($\langle \Delta R\rangle \simeq 0.02$) with those
  done using Monte Carlo simulations, but make much simpler
  assumptions.  Further, for about half the systems, the observed flux
  anomaly is very similar to those estimated by both groups.}
\label{fg:compare}
\end{figure}

In Fig.~\ref{fg:compare}, we compare our estimates of flux ratio
anomalies with those found using Monte Carlo analysis by KGP, and
those observed. A detailed description of each system (including three
not analyzed by KGP) follows.  It is noteworthy that most of our
reconstructions produces similar estimates to flux anomalies as those
found by KGP.  However, many systems exhibit fold relations
inconsistent with smooth lensing models.  For lobe-dominated radio sources,
this likely suggests substructure within the lens.  For
core-dominated sources, microlensing may be at work, depending on the
beaming of the radio lobe.  In the latter case, this ambiguity could
presumably be resolved by exploring the lenses in the time domain.\\

\noindent {\it B0128+437} is a system with both an extremely high flux
anomaly, as seen at several different frequencies
\citep{Biggs/Browne:2004,Phillips/Norbury:2000}, as well as multiple
components in the source.  Images A, C and D have been resolved into
three different components embedded into a more extended jet.  Since
image B's components are not well-defined, we can only directly
consider the integrated flux ratio here \citep{Biggs/Browne:2004}.
Moreover, since the observed components of images A and B are clearly
resolvable, shape analysis naturally would yield additional
information \citep{Koopmans/etal:2003}.

Observationally, this system exhibits a very large flux anomaly in the
range of $R=0.263$ \citep{Koopmans/etal:2003} to $R=0.582$
\citep{Biggs/Browne:2004}.  KGP have studied this system and found
from simulations an expected flux anomaly in the range of $-0.1$ to
$0.4$, with a preferred value around $R=0.25$.  Our own analysis of
this system predicts a value of $R=0.161$ for an SIS+ES and $R=0.176$
for a PS+ES, well within the range predicted by simulation.  Further,
high-resolution imaging of the system resolves unambiguous lobes,
suggesting a source too large to be affected by microlensing.  This
points unambiguously to additional substructure in the lensing galaxy.
We will explore the question further constraining substructure in
future work.\\

\noindent{\it B0712+472} is classified by KGP as both a cusp and a
fold. B0712+472 violates the cusp relation, as discovered in Keeton et
al. (2003), but only in optical data, which is evidence more for
microlensing than for small-scale structure. The four images are
relatively circular and easily distinguishable from each other, but
there is a faint "bridge-like'' feature between components A and B,
which are very close together \citep{Jackson/Nair:1998}, suggesting a
slightly extended source at the highest resolution.
Spectroscopically, the quasar has a relatively flat spectrum which,
along with most photometric estimates, imply a core-dominated source.

The KGP Monte Carlo simulations predict $R_{fold}$ to be between -0.1
and 0.15. Our value for $R_{fold}$ from fitting the lens to an SIS+ES
is -0.023, in agreement with the KGP simulations. The observed radio
observations
\citep{Jackson/Nair:1998,Jackson/Xanthopoulos:2000,Koopmans/etal:2003}
also agree with the predictions.\\

\noindent{\it B1555+375} images A and B appear in radio observations
as one extended object with two separate peaks of brightness. The
fourth image was not initially observed but was predicted, searched
for and consequently discovered \citep{Marlow/Myers:1999}. KGP find an
expected range for the value of $R_{fold}$ for this lens to be from
about -0.07 to 0.2. Our model using an SIS+ES produces an $R_{fold}$
of about 0.003, which goes up to 0.062 if we use a PS+ES. The observed
$R_{fold}$, which is greater than 0.23 in all radio observations,
exceeds our estimate, suggesting the system has significant
small-scale structure or potentially microlensing.  The imaging has
insufficient resolution in the radio to determine whether the source
is extended.  However, no double-lobes or jets are clearly visible.\\

\noindent{\it B1608+656} has a core-dominated source with a complex
lensing structure in the form of a second
galaxy. \cite{Myers/Fassnacht:1995} claim that B1608+656 is simple
compared to other lens systems because they were able to reproduce the
image configuration using a single, elliptical lens galaxy in their
model. The ellipticity of the lens in their models turns out to have
been a sufficiently accurate recreation of the combination of lensing
galaxies in the real system.

KGP predicts $R_{fold}$ to be in an unusually small range, from 0.3 to
0.43.  Observed $R_{fold}$ values, which range from 0.327 to 0.516,
match up well \citep{Myers/Fassnacht:1995, Fassnacht/Pearson:1999,
  Snellen/deBruyn:1995, Fassnacht/Xanthopoulos:2002}. Our estimate for
a SIS+ES is $R_{fold}=0.361$, and for PS+ES is $R_{fold}=0.577$, which
is higher than both observations and other predictions, suggesting a
more isothermal mass distribution. \\

\noindent {\it B1933+503} has ten images as a result of three lensed
images.  Two of the sources result in quad formations and the third
appears doubly imaged.  As noted by \cite{saasfees}, the three sources
seem to correspond to a core and both radio lobes. The four images
that appear in a standard quad formation are a cross-like fold image
set.

KGP model $R_{fold}$ in a rather wide range from about -0.1 to 0.45,
with a sharp peak at 0.05 and a shorter, wider peak at 0.3. The
observed flux ratios (at any wavelength) are much higher: in the range
of 0.580 to 0.722 \citep{Sykes/Browne:1998}.  Our geometric estimate
of the flux ratio using a standard SIS+ES is $-0.233$. This anomalous
outlier presents a limit to our approach, since \cite{Nair:1998}
model with a lens ellipticity of 0.81.\\

\noindent{\it B1938+666} is a lens with two sources, quite possibly
extended radio lobes.  The radio lobe-dominated emission picture is
further supported by a spectral index of about 0.5.  One source
creates a four image fold configuration, which we use for further
analyses. The other source produces a double
~\citep{King/Browne:1997}. In the near-infrared and optical wavebands,
a slight Einstein ring appears ~\citep{King/Jackson:1998}.

The two fold images have a very small flux anomaly as detected using
MERLIN, but a much higher value, $-0.436$, using VLBI
~\citep{King/Browne:1997}.  A geometric reconstruction of this lens
is much closer to the high value, which either suggests substructure
or perhaps significant time variability.  The observed images are
resolved enough to measure shapes, which would potentially further
constrain mass models.

\noindent{\it HS0810+2554} has not been observed in the radio, but we
used the near-infrared CASTLES observations which yield a flux ratio
anomaly of 0.274. The lens was discovered
by \cite{Reimers/Hagen:2002}, who noted that this system's
configuration is very similar to that of PG1115+080. Our geometric
analysis yields a value much closer to zero, however, suggesting
either small-scale structure, or possibly differential absorption.\\

\noindent {\it MG0414+0534} has been observed over a wide range of
frequencies and epochs. From optical observations, there is a
non-distinct arc that passes through the fold image pair, A1 and A2,
and the next closest image, B ~\citep{Angonin/Vanderriest:1999}. Radio
data clearly shows the resolution of the images in this fold
configuration lens system ~\citep{Katz/Hewitt:1993}.

MG0414+0534 has lobe-dominated emission and an extremely steep radio
spectrum.  It also has a satellite galaxy, ``object X'', that is not
particularly close to the fold image pair. In order to match the image
positions, KGP needed to include object X in the model. The $R_{fold}$
values estimated from KGP range from about 0.0 to 0.2 with a peak at
around 0.07. Observed data has $R_{fold}$ values ranging from about
0.047 to 0.073 (\citealt{Katz/Moore:1997, Katz/Hewitt:1993,
  Hewitt/Turner:1992} and CLASS) which agrees our estimates.  We did
not include object X in our analysis, which resulted in a somewhat
lower flux
anomaly ration than observed or estimated by previous analysis.\\

\noindent{\it MG2016+112} has two separate sources, lensing to a quad
and a double, to produce a total of six images
\citep{King/Browne:1997}.  Not all image components are visible in the
various wavelengths.  ~\cite{Schneider/Lawrence:1985} observed three
images and one lens galaxy in both radio and optical observations. One
year later, ~\cite{Schneider/Gunn:1986} found an additional
image. CASTLES has observed three images and four lens galaxies.

We reconstructed this lens with a lens position from the CASTLES
optical data, and image positions from ~\cite{More/McKean:2009}'s
1.7GHz data.  They conclude, ``there is no significant substructure or
any other effects that might affect the flux densities of the
images.''  Our reconstruction yields a flux fold ratio
consistent to that observed, further supporting
\cite{More/McKean:2009}.
\\

\noindent{\it PG1115+080} has thus far been measured only in the
mid-IR. The NICMOS images of PG1115+080 suggest a quasar host galaxy
that lenses to an Einstein ring around the four fold configuration
images. The lensing galaxy has no substructure, and there is no
blatant flux anomaly in the infrared ~\citep{Impey/Falco:1998}.

KGP predicts the value of $R_{fold}$ to range from -0.05 to 0.25, with
a peak at about 0.1. This is consistent with observed values
~\citep{Chiba/Minezaki:2005,Impey/Falco:1998,Chiba/Minezaki:2005}.
Our own reconstructed fold relation is comfortably in this range
as well.

From our reconstruction of the image
positions using an SIS+ES yield an $R_{fold}$ value of about 0.045
which fits our observed infrared data from ~\cite{Chiba/Minezaki:2005}
and fits perfectly in the range determined by KGP. This further
supports the thought that this lens has little to no substructure. \\ 

\noindent{\it SDSS1004+4112} is a very massive system
\citep{Inada/Oguri:2003} which has only thus far been reliably
measured in the near-IR.  It contains a five-image fold-configuration
system with a complicated galaxy cluster at the heart of its lens
\citep{Inada/Oguri:2005}.

KGP predicts SDSS1004+4112 to have $R_{fold}$ values between 0 and
0.25, with a sharp peak at 0, with higher flux ratios less likely.
Observations fall toward the high end of this range, as does our
own geometric reconstruction.  This system is known to have complex
lens structure, but no flux ratio anomaly occurs in the fold pair
images. \\

\noindent{\it SDSS1330+1810} is an apparently lobe-dominated source
observed in the near-infrared and optical wavebands with the Magellan,
UH88, and ARC3.5m telescopes. The fold relation is vastly different in
the optical and near-infrared, thus ~\cite{Oguri/Inada:2008} predicts
dust to be the cause of this anomaly. From analyzing the images, there
may be some substructure in the form of a cluster of galaxies
positioned near the lens ~\citep{Oguri/Inada:2008}.

We performed a reconstruction of this lens with only the near infrared
data in the $J$, $H$, and $K_s$ bands. The values of $R_{fold}$ from
observed near-infrared data ranged from 0.101 to 0.151. Our geometric
reconstruction predicts a smaller fold relations, of only 0.007,
suggesting contribution of substructure.

\section{Conclusions and Future Work}

\label{sec:future}

Thus far, we have focused on estimating the flux anomaly for fold
lenses using entirely geometric arguments.  This is approach is useful
because it produces robust estimates of flux fold relations without the
need to use Monte Carlo simulations. In fact, the only model-specific,
non-observed, parameter to be adjusted is the index of the radial
profile which can be tuned analytically.  This is both easier to apply
than past approaches, and also provides a much more intuitive
understanding of the underlying structure of the lens.  However, it is
worth noting that Monte Carlo analysis is still necessary to get a
deeper understanding of the underlying variance in the fold relation
distribution.  

This work is an important first step in further geometric,
semi-analytic analysis methods, and has a number of potential future
offshoots.  For one, following \cite{Congdon:2008}, we will extend our
analysis to include analytic estimates of time delays.  Since time
delays are functions of only the potential itself, the analysis is
expected to be significantly more robust.  Further, in
\S\ref{sec:substructure}, we performed a tentative analysis of a
system with substructure.  Future work will be required to identify
the degenerate families of substructure which can identified by
geometric analysis.  Another logical next step is to perform a
geometric analysis of cusp lenses \citep{Keeton/Gaudi:2003}.  The
central focus of this work, however, is based on the premise of
uniquely identifying substructure in lenses and our long term focus
will be to provide a systematic analysis of substructure in quad
lenses using this framework.

\section*{Acknowledgments}

The authors would like to gratefully acknowledge useful conversations
with C. Keeton, and some helpful notes on the manuscript from
R. Kratzer.  We also thank the anonymous referee for extensive
comments which significantly improved the final manuscript. This work
was supported by NSF Award 0908307, and the Drexel University,
``Students Tackling Advanced Research'' program.

\bibliography{goldberg_cites.bib}

\newpage

\begin{longtable}{| c | c | c || c | c | c | c |}
\hline Lens & $\theta_E$ & $R_{fold}$
(geom.) & $R_{fold}$ (obs.) & $d_1$ & Frequency & Reference \\
\hline

B0128+437 & 0.200 & 0.161 & 0.331 $\pm$ 0.118 & 0.186 & MERLIN 5GHz & \cite{Phillips/Norbury:2000}\\
& & & 0.263 $\pm$ 0.017 & & MERLIN 5GHz & \cite{Koopmans/etal:2003}\\
& & & 0.212 & & VLBA 5GHz & \cite{Biggs/Browne:2004}\\
\hline

B0712+472 & & -0.023 & 0.041 $\pm$ 0.105 & 0.168 & MERLIN 5GHz & \cite{Jackson/Nair:1998}\\
& & & 0.105 $\pm$ 0.103 & 0.168 & MERLIN 5GHz & \cite{Jackson/Nair:1998}\\
& & & 0.097 $\pm$ 0.069 & 0.168 & VLBA 5GHz & \cite{Jackson/Nair:1998}\\
& & & 0.147 $\pm$ 0.102 & 0.168 & VLA 15GHz & \cite{Jackson/Nair:1998}\\
& & & 0.105 $\pm$ 0.031 & 0.168 & HST 814nm & \cite{Jackson/Nair:1998}\\
& 0.680 & & 0.085 $\pm$ 0.036 & 0.170 & MERLIN 5GHz & \cite{Koopmans/etal:2003}\\

\hline

B1555+375 & & 0.003 & 0.273 $\pm$ 0.124 & 0.087 & MERLIN 5GHz & \cite{Marlow/Myers:1999}\\
& & & 0.280 $\pm$ 0.123 & 0.087 & VLA 15GHz & \cite{Marlow/Myers:1999}\\
& & & 0.235 $\pm$ 0.022 & & MERLIN 5GHz &\cite{Koopmans/etal:2003}\\
\hline

B1608+656 & 0.770 & 0.361 & 0.416 $\pm$ 0.019 & 0.880 & VLA 8.4GHz & \cite{Myers/Fassnacht:1995}\\
& & & 0.516 $\pm$ 0.058 & 0.880 & VLA 8.4GHz & \cite{Snellen/deBruyn:1995}\\
& & & 0.402 $\pm$ 0.055 & 0.880 & VLA 15GHz & \cite{Snellen/deBruyn:1995}\\
& & & 0.327 & 0.872 & VLA 8.4GHz & \cite{Fassnacht/Pearson:1999}\\
& & & 0.346 & 0.872 & VLA 8.5GHz & \cite{Fassnacht/Xanthopoulos:2002}\\
& & & 0.326 & 0.872 & VLA 8.5GHz & \cite{Fassnacht/Xanthopoulos:2002}\\
\hline

B1933+503 & 0.49 & -0.233 & 0.580 $\pm$ 0.048 & 0.457 & MERLIN 1.7GHz &\cite{Sykes/Browne:1998}\\
& & & 0.610 $\pm$ 0.023 & 0.457 & VLBA 5GHz & \cite{Sykes/Browne:1998}\\
& & & 0.644 $\pm$ 0.030 & 0.457 & VLA 8.4GHz & \cite{Sykes/Browne:1998}\\
& & & 0.722 $\pm$ 0.031 & 0.457 & VLA 15GHz & \cite{Sykes/Browne:1998}\\
& & & 0.968 $\pm$ 0.041 & & VLBA 1.7GHz & \cite{Marlow/Browne:1999}\\
& & & 0.974 $\pm$ 0.039 & & VLBA 1.7GHz & \cite{Marlow/Browne:1999}\\
& & & 0.668 $\pm$ 0.027 & & VLA 8.4GHz & \cite{Biggs/Xanthopoulos:2000}\\
& & & 0.644 $\pm$ 0.030 & & VLA 8.4GHz & \cite{Biggs/Xanthopoulos:2000}\\
& & & 0.653 $\pm$ 0.030 & & VLA 8.4GHz & \cite{Biggs/Xanthopoulos:2000}\\
\hline

B1938+666 & & -0.573 & -0.0103 & 0.147 & MERLIN 5GHz & \cite{King/Browne:1997}\\
& & & -0.047 & & MERLIN 1.612GHz & \cite{King/Browne:1997}\\
& & & -0.436 & & VLBI 1.7GHz & \cite{King/Browne:1997}\\
\hline

HS0810+2554 & & 0.003 & 0.274 $\pm$ 0.009 & 0.185 & HST F160W & CASTLES\\
\hline

MG0414+0534 & 1.08 & -0.014 & 0.054 $\pm$ 0.003 & & VLA 8 GHz & \cite{Katz/Moore:1997}\\
& & & 0.054 $\pm$ 0.006 & & VLA 8 GHz & \cite{Katz/Moore:1997}\\
& & & 0.051 $\pm$ 0.015 & & VLA 5 GHz & \cite{Katz/Moore:1997}\\
& & & 0.066 $\pm$ 0.028 & & VLA 15GHz & \cite{Katz/Moore:1997}\\
& & & 0.060 $\pm$ 0.027 & & VLA 15GHz & \cite{Katz/Moore:1997}\\
& & & 0.063 $\pm$ 0.018 & & VLA 15GHz & \cite{Katz/Moore:1997}\\
& & & 0.058 $\pm$ 0.028 & & VLA 15GHz & \cite{Katz/Moore:1997}\\
& & & 0.064 $\pm$ 0.035 & 0.426 & VLA 22GHz & \cite{Katz/Moore:1997}\\
& & & 0.054 $\pm$ 0.006 & 0.412 & VLA 8GHz & \cite{Katz/Hewitt:1993}\\
& & & 0.073 $\pm$ 0.015 & & VLA 15GHz & \cite{Katz/Hewitt:1993}\\
& & & 0.672 $\pm$ 0.008 & & VLA 5 GHz & \cite{Hewitt/Turner:1992}\\
& & & 0.666 $\pm$ 0.007 & & VLA 15GHz & \cite{Hewitt/Turner:1992}\\
& & & 0.047 & 0.409 & VLA 8.4GHz & CLASS \\
\hline

MG2016+112 & 1.570 & 0.169 & 0.205$\pm$ 0.014 & 0.0428 & MERLIN 5GHz& \cite{More/McKean:2009}\\
\hline

PG1115+080 &1.030 & 0.045 & 0.036$\pm$ 0.024 & 0.482 & COMICS on Subaru Telescope & \cite{Chiba/Minezaki:2005}\\
& & & 0.218 $\pm$ 0.012 & 0.485 & HST/NICMOS F160W & \cite{Impey/Falco:1998}\\
& & & 0.226 $\pm$ 0.009& 0.482 & HST/NICMOS F160W & CASTLES \\
\hline

SDSS1004+4112 & 6.910 & 0.174 & 0.213 $\pm$ 0.016 & 3.767 & HST F160W & CASTLES\\
& & & 0.155 $\pm$ 0.068 & 3.770 & HST NICMOS & \cite{Inada/Oguri:2005}\\
\hline

SDSS J1330+1810& & 0.007 & 0.146 $\pm$ 0.029 & 0.420 $\pm$ 0.004 & Magellan J & ~\cite{Oguri/Inada:2008} \\
 & & & 0.101 $\pm$ 0.027 & & ARC2.5m H & ~\cite{Oguri/Inada:2008} \\
 & & & 0.151 $\pm$ 0.040 & & Magellan H & ~\cite{Oguri/Inada:2008} \\
 & & & 0.110 $\pm$ 0.022 & & Magellan $K_s$ & ~\cite{Oguri/Inada:2008} \\ 
\hline

\caption{Observed IR and radio fold systems. For systems observed at
multiple wavelengths, positional information was used in that band
with the lowest error in image position. Galaxy position was
typically measured in the visible. Errorbars are listed for all
observations or papers where errors are reported. All distances and
Einstein radii are in arcseconds. Flux ratios are dimensionless.
Estimates of the Einstein radius are taken from KGP. The
geometrically modeled flux ratio anomalies are fit using a Singular
Isothermal Sphere plus external shear profile.}
\label{tab:data}
\end{longtable}

\end{document}